\documentclass[a4paper,11pt]{article}
\usepackage{pos}
\renewcommand{\logo}{\relax}
\renewcommand{\hookBeforeTitle}{\hfill\textsc{BU-HEPP-22-03}
        \vskip2em   plus .4fill  minus 1em}

\title{KKMChh: matching CEEX photonic ISR to a QED-corrected parton shower}
\author*[a]{S. A. Yost}
\author[b]{S. Jadach}
\author[c]{B. F. L. Ward}
\author[b]{Z. W{\c a}s}

\affiliation[a]{Department of Physics, The Citadel,\\
  171 Moultrie St., Charleston, South Carolina 29409, USA}

\affiliation[b]{Institute of Nuclear Physics, Polish Academy of Science,\\
ul. Radzikowskiego 152, 31-342 Krak\'{o}w, Poland}

\affiliation[c]{Department of Physics, Baylor University,\\
One Bear Place 97316, Waco, Texas 76798, USA}

\emailAdd{Scott.Yost@citadel.edu}
\emailAdd{Stanislaw.Jadach@cern.ch}
\emailAdd{BFL\_Ward@baylor.edu}
\emailAdd{Z.Was@cern.ch}

\abstract{KKMChh adapts the CEEX (Coherent Exclusive Exponentiation) formalism
of the Monte Carlo Program KKMC for Z boson production and decay to hadron 
scattering. Amplitude-level soft photon exponentiation of initial and final 
state radiation, together with initial-final interference, is matched to a 
perturbative calculation to second order next-to-leading logarithm, and 
electroweak corrections to the hard process are included via DIZET. The first 
release of KKMChh included complete initial state radiation 
calculated with current quark masses. This version assumes idealized pure-QCD 
PDFs with negligible QED contamination. Traditional PDFs neglect QED evolution
but are not necessarily free of QED influence in the data. QED-corrected PDFs 
provide a firmer starting point for precision QED work. We describe a new 
procedure for matching KKMChh's initial state radiation to a QED-corrected 
PDF, and compare this to earlier approaches.}

\FullConference{%
  41st International Conference on High Energy physics - ICHEP2022\\
  6-13 July, 2022\\
  Bologna, Italy
}

\begin{document}
\maketitle

\section{Introduction}

KKMChh~\cite{kkmchh1,kkmchh2} adapts Coherent Exclusive 
Exponentiation (CEEX)~\cite{ceex} to hadronic collisions producing lepton
pairs with multi-photon radiation,  $pp\to Z/\gamma^*\to l{\bar l}+n\gamma$.
CEEX is an amplitude-level implementation of YFS exponentiation~\cite{yfs}.
KKMChh includes hard photon residuals through order $\alpha^2 L$,
%\cite{alpha2L}, 
where $L$ is an appropriate ``big logarithm.'' 
A separate DIZET6.45 library~\cite{dizet1} tabulates 
electroweak form factors before a MC run.
KKMC~\cite{kkmc} was originally designed for $e^+e^-$ collisions. The 
was recently reprogrammed in C++ and released as KKMCee 5.00.2~\cite{kkmcee}.
Here, we focus on the effect of ISR on the quark joint parton luminosity
 and describe ``Negative ISR,'' a recent addition to KKMChh to allow it to be 
used with parton distribution functions (PDFs) incorporating QED corrections. 

\section{Joint Parton Luminosity}

We will examine the effect of ISR radiation on the colliding quarks by 
focusing on the joint parton luminosity function for a quark $q$.  Consider a
collision between quarks with momentum fractions $x_q$, $x_{\bar q}$ in a 
proton collision with CM energy $\sqrt{s}$. Neglecting quark masses relative 
to the CM energy, the squared CM energy of the quark collision is ${\hat x}s$ 
with ${\hat x}\equiv x_q x_{\bar q}$. Without QED corrections, the cross 
section for $pp\to Z/\gamma^*\to f{\bar f}$ would be proportional to 
the joint parton luminosity $L_{q{\bar q}}$ defined in terms of PDFs 
$f_q$, $f_{\bar q}$ by 
\begin{equation}
\label{Ldist}
L_{q{\bar q}}({\hat x}s) = 
\int dx_q dx_{\bar q}\delta({\hat x} - x_q x_{\bar q})
f_q(x_q,{\hat x}s) f_{\bar q}(x_{\bar q},{\hat x}s).
\end{equation}

Exponentiation of photonic ISR in a soft-photon approximation leads to an 
ISR radiation factor of the form~\cite{kkmc,yfs}
\begin{equation}
\rho_{\rm ISR}^{(0)}(v, \gamma) = F_{\rm YFS}(\gamma)\gamma v^{\gamma-1}\ ,
\end{equation} 
where $v$ is the soft photon radiation fraction, leaving the quarks with 
squared CM energy $xs = (1-v){\hat x}s$, and 
\begin{equation}
F_{\rm YFS}(\gamma) = \frac{e^{-C_E \gamma}}{\Gamma(1+\gamma)}
\end{equation}
is the YFS form factor, $C_E\approx 0.57722$ is Euler's constant, and 
\begin{equation}
\gamma \equiv \gamma_{\rm ISR}({\hat x}s,Q_q, m_q) = \frac{2\alpha}{\pi}Q_q^2
 \left[ \ln\left(\frac{{\hat x}s}{m_q^2}\right) - 1\right]\ .
\end{equation} 
We use current quark masses $m_d = 4.7$\ MeV, $m_u = 2.2$\ MeV, 
$m_s = 93$\ MeV, $m_c = 1.2$\ GeV, and $m_b = 4.7$\ GeV.~\cite{pdg} 
The superscript $(0)$ indicates that this ISR radiation factor includes 
no fixed-order corrections, and is an approximation to the full 
$\rho_{\rm ISR}^{(2)}$ in KKMChh, which has fixed-order corrections through 
$\alpha^2 L$, where $L = \ln({\hat x}s/m_q^2)$.

\begin{figure}[h]
\setlength{\unitlength}{\textwidth}
\begin{picture}(1,0.35) 
\put(0,0){\includegraphics[width=0.48\textwidth]{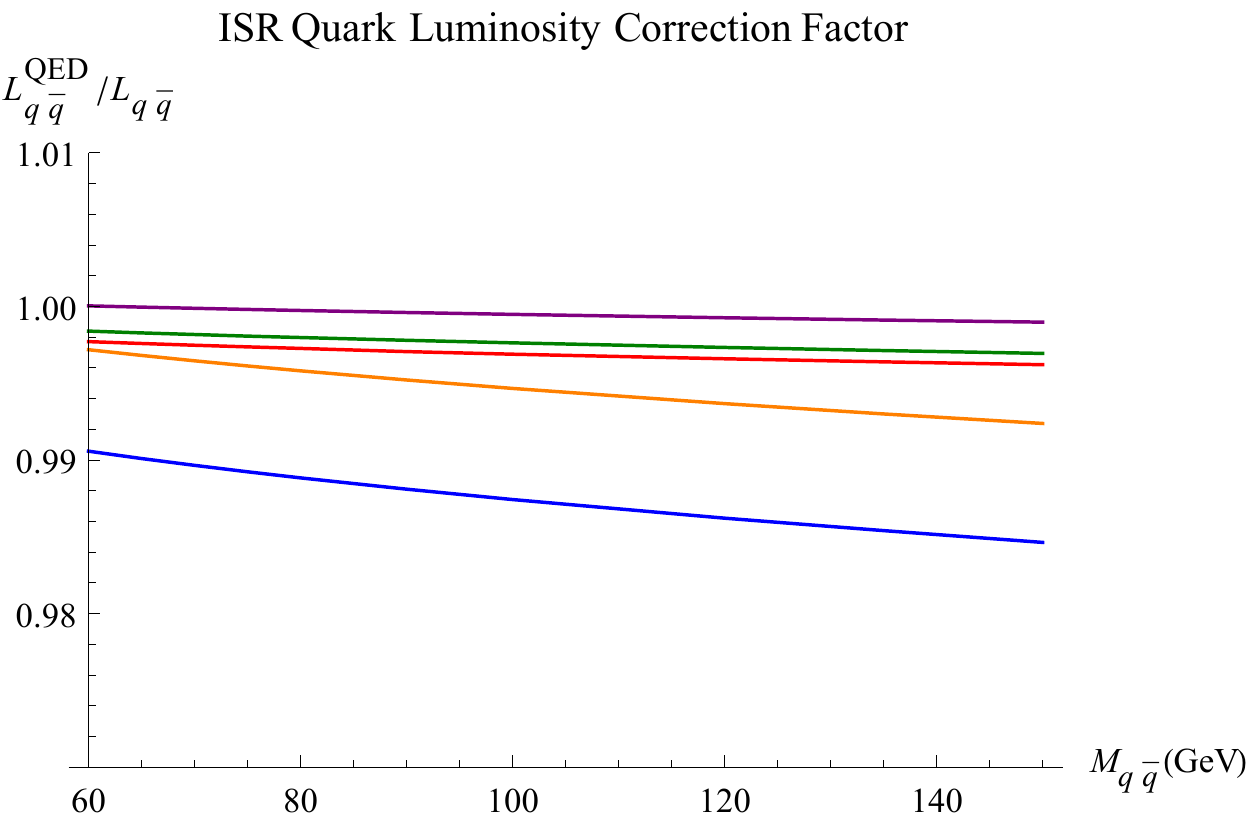}}
\put(0.48,0){\includegraphics[width=0.48\textwidth]{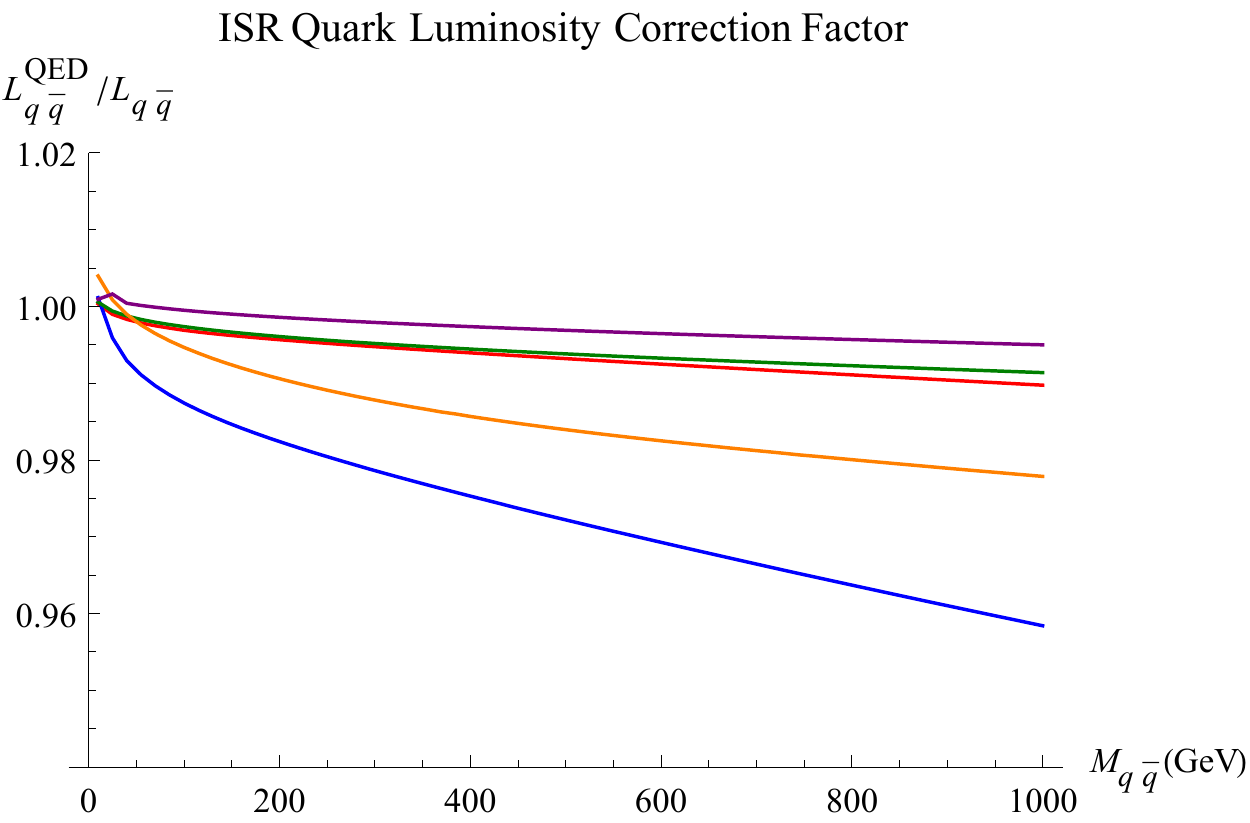}}
\put(0.9,0.1){\includegraphics[width=0.075\textwidth]{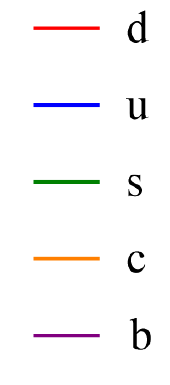}}
\end{picture}
\caption{These graphs show the ratios of joint quark luminosity distributions
$L_{q{\bar q}}^{\rm QED}/L_{q{\bar q}}$ with and without photonic ISR
for each quark in two ranges: $60 - 150$\ GeV on the left and $10 - 1000$\ GeV 
on the right. The graphs were produced using ManeParse~\cite{maneparse} with 
NNPDF3.1 NLO PDFs~\cite{nnpdf1}.}
\label{QuarkLum}
\end{figure}

The joint quark luminosity function after photonic ISR is then 
\begin{equation}
\label{LdistQED}
%xL^{\rm QED}_{q{\bar q}}(xs) = \int_0^{1-x}dv\int_x^1 d{\hat x} 
%  \delta(x - (1-v){\hat x})\;{\hat x}
%  L_{q{\bar q}}({\hat x}s)\;(1-v)\rho_{\rm ISR}^{(0)}(v, \gamma_{\rm ISR}(
%	{\hat x}s,Q_q,m_q)).
L^{\rm QED}_{q{\bar q}}(xs) = \int_0^{1-x}dv 
  L_{q{\bar q}}\left(\frac{xs}{1-v}\right)\;
  \rho_{\rm ISR}^{(0)}\left(v, \gamma_{\rm ISR}
  \left(\frac{xs}{1-v},Q_q,m_q\right)\right).
\end{equation}
Fig.\ \ref{QuarkLum} shows the ratio of eq.\ \ref{LdistQED} to 
eq.\ \ref{Ldist} for each quark, using NNPDF3.1 NLO PDFs.~\cite{nnpdf1}
Over the narrower range on the left, the ratio varies slowly, and at  
$M_{q{\bar q}} = M_Z$, QED ISR reduces the joint luminosity distribution
by $-1.2\%$ for the up quark and $-0.30\%$ for the down quark. 
Over the range $10 - 1000$\ GeV, the up quark 
distribution varies from $+0.11\%$ to $-4.2\%$, while the down quark 
distribution varies less, from $+0.30\%$ to $-1.0\%$.

\section{PDF Matching via Negative Initial State Radiation}

In its original form, KKMChh was intended to be used with PDF sets having a
negligible photonic contribution, adding complete \emph{ab initio} QED ISR
to idealized PFDs representing pure QCD. Even without QED evolution, QED can 
influence a PDF via the data.
It is conceptually cleaner to start with a PDF set that incorporates QED 
corrections consistently. To match KKMChh's exponentiated ISR to a 
QED-corrected PDF set without double-counting, we introduced a feature 
called ``Negative ISR,'' or NISR, which backs out QED from the PDF using ISR 
radiator factors in reverse, starting from a
scale $q_0$ which, for QED-corrected PDFs, we take to be the hard process
scale $sx$. 

In short, the idea is to combine each
PDF function with inverse ``half-radiator'' factors
\begin{equation}
\rho_{\rm ISR}^{(2)}\left(u_q, 
	-\frac12\gamma_{\rm ISR}({\hat x}s)\right), 
\quad\rho_{\rm ISR}^{(2)}\left(u_{\bar q},
	-\frac12\gamma_{\rm ISR}({\hat x}s)\right)  
\end{equation}
and define modified quark momentum fractions $x'_q = x_q(1-u_q)$, 
$x'_{\bar q} = x_{\bar q}(1-u_{\bar q})$ before ISR, so that when the PDFs
are convoluted with these inverse half-radiators and the forward KKMChh
radiator 
$\rho_{\rm ISR}^{(2)}\left(v,\gamma_{\rm ISR}\left({\hat x}s\right)\right)$, 
the original 
joint luminosity distribution is recovered, and ISR photons are generated
with a modified momentum fraction $v'$ satisfying the constraint
\begin{equation}
1 - v' = (1-v)(1-u_q)(1-u_{\bar q})\ .
\end{equation}

When we calculated the inclusive quark-level Drell-Yan cross section for muons 
with 60 GeV $< M_{\mu\mu}<$ 150 GeV using NNPDF3.1 NLO alone and then 
turned on KKMChh ISR with NISR, we found that the original PDF-only result is 
recovered to within $0.3\%$ for charge $\frac23 e$ quarks and $0.08\%$ for 
charge $-\frac13 e$ quarks. Both of these differences are due to an 
${\cal O}(\alpha)$ non-logarithmic correction
\begin{equation}
\delta_Q = Q^2 \frac{\alpha}{\pi}\left(-\frac12 + \frac{\pi^2}{3}\right).
\end{equation} 
This term is not included in NISR evolution and precisely accounts for the
difference seen. We also found that the quark-level cross sections with ISR
and NISR are unchanged if the quark masses are all set to a high value of 
500 MeV, demonstrating the quark mass independence of this procedure.

\begin{figure}[b]
\setlength{\unitlength}{\textwidth}
\begin{picture}(1,0.4) 
\put(0,0){\includegraphics[width=\textwidth,height=0.4\textwidth]{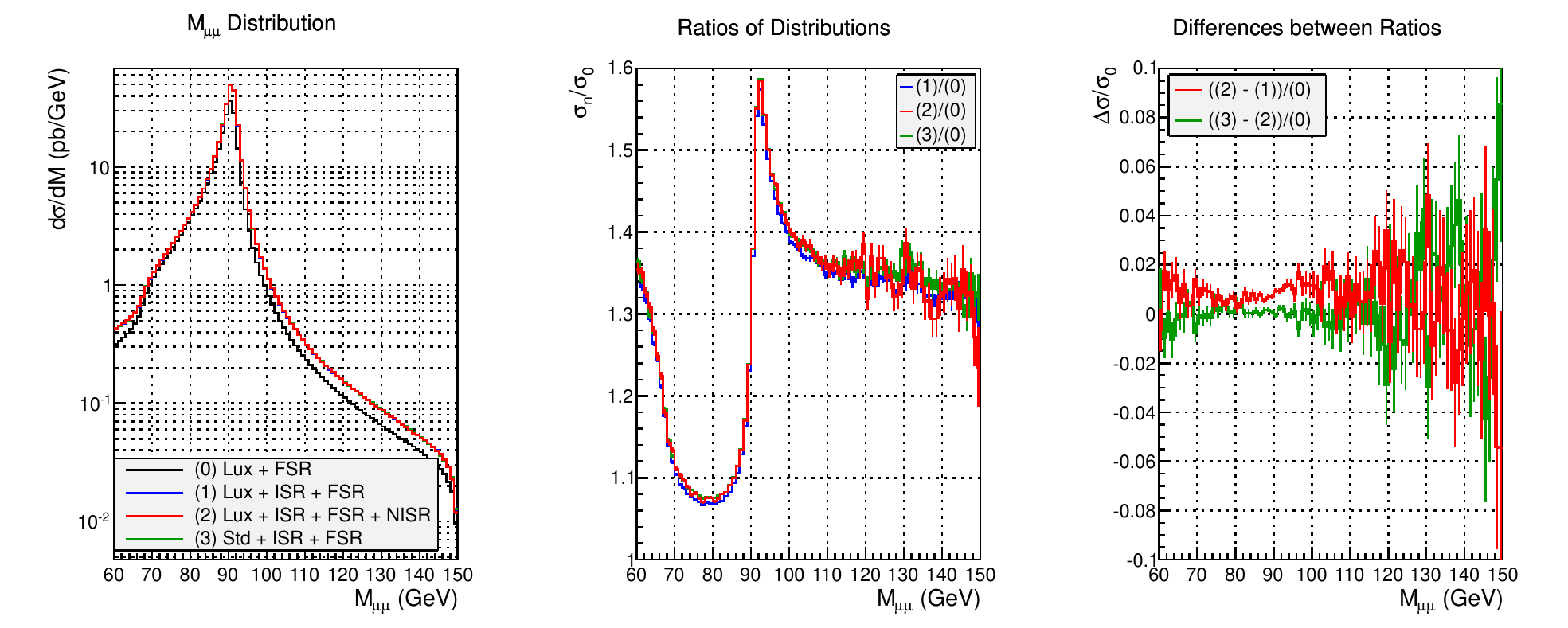}}
\end{picture}
\caption{The invariant mass $M_{\mu\mu}$ distribution of the final muon
pair for events with at least one photon and each muon having 
$p_{\mathrm{T}\mu} > 25$ GeV, $\eta_\mu < 2.5$ calculated with (0) FSR only 
(black), (1) FSR + ISR (blue), and (2) FSR + ISR with NISR (red) for 
NNPDF3.1-LuxQED NLO PDFs. For comparison, (3) shows FSR + ISR with ordinary 
NNPDF3.1 NLO PDFs (green). The center graph shows ISR on/off ratios $(1)/(0)$ 
(blue), $(2)/(0)$ (red) and $(3)/(0)$  (green).  The right-hand graph shows 
the fractional differences $((1) - (2))/(0)$ in red and $((2) - (3))/(0)$ in 
green. }
\label{Mll}
\end{figure}

\begin{figure}[t]
\setlength{\unitlength}{\textwidth}
\begin{picture}(1,0.4) 
\put(0,0){\includegraphics[width=\textwidth,height=0.4\textwidth]{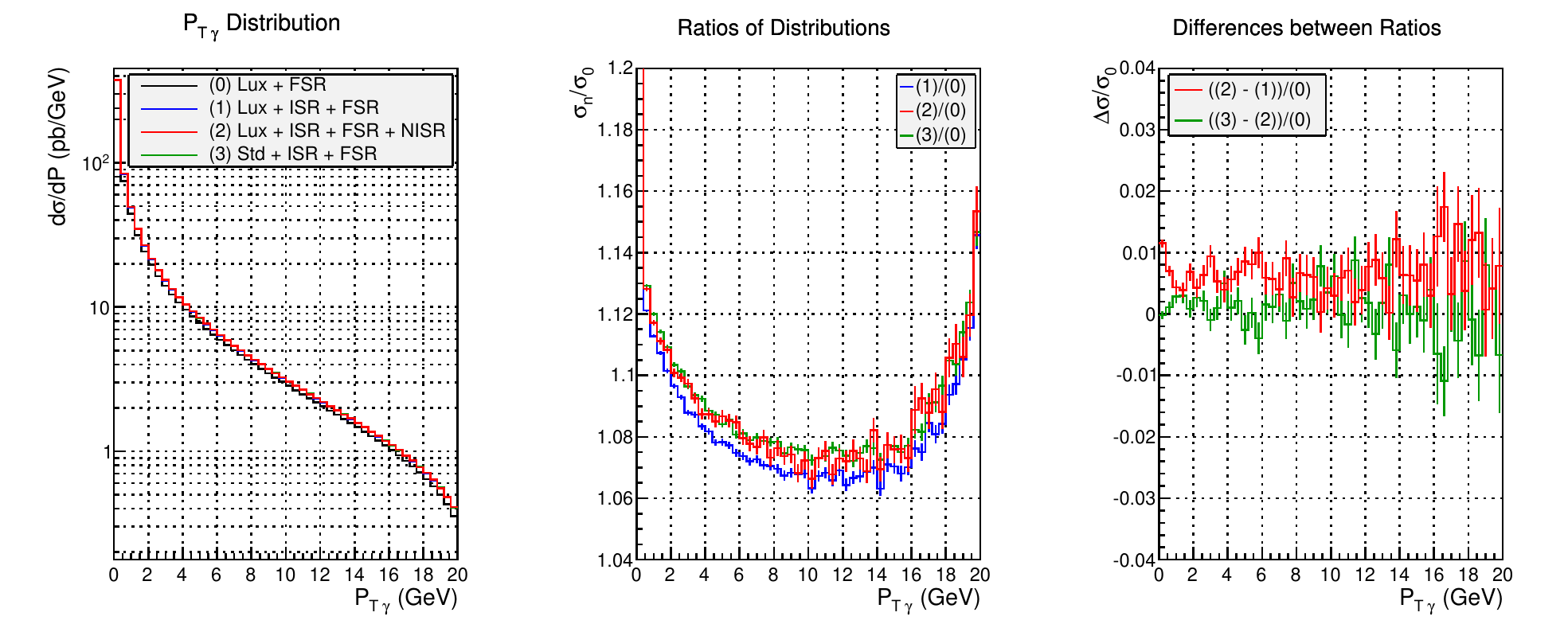}}
\end{picture}
\caption{The transverse momentum distribution for $P_{\mathrm{T}\gamma}$ of 
the photon 
for which it is greatest in the same event set as Fig.\ \ref{Mll}, showing the 
same comparisons among the various ways of adding ISR. }
\label{PTg}
\end{figure}

Figs.\ \ref{Mll} and \ref{PTg} show the dimuon invariant mass and transverse
momentum of the hardest photon for approximately 200 million KKMChh events
including at least one photon, with a requirement that each muon have 
transverse momentum $p_{\mathrm{T}\mu} > 25$ GeV and pseudorapidity 
$|\eta_\mu| < 2.5$.
All results except (3) (green) use NNPDF3.1-LuxQED NLO PDFs~\cite{nnpdf2}, 
while (3) uses standard NNPDF3.1 NLO PDFs~\cite{nnpdf1}. The black histogram 
(0) on the left side of each figure includes FSR only with quarks generated 
using a LuxQED PDF set.  The blue histogram (1) adds 
ISR to quarks generated using the LuxQED version without NISR, the red 
histogram (2) does the same with NISR, and the green histogram (3) adds ISR 
to quarks generated using the ordinary PDFs. 

The center graph in each figure shows the fractional ISR contribution for 
cases (1), (2) and (3) in blue, red, and green, respectively.  The red plot in 
the graph on the right of each figure is the difference between the 
fractional ISR contributions with and without NISR. We see that NISR makes a 
change between 0.5\% and 1.0\% in each distribution. The green plot shows the 
difference between the fractional ISR contribution calculated with an ordinary
PDF set without NISR and the fractional ISR contribution calculated with a
LuxQED PDF set including NISR. For NNPDF3.1, these two methods agree within 
statistical errors. 

The agreement shown by the green plots on the right of each figure can be 
traced to a close agreement between the ratio 
$L_{q{\bar q}}^{\rm QED}/L_{q{\bar q}}$ of joint quark luminosity distributions
with ISR on/off in Fig.\ \ref{QuarkLum} and the ratio of joint quark luminosity
distributions calculated for NNPDF with and without LuxQED at scales near 
the $Z$ mass. 
%NISR adds considerable computational overhead, since its primary MC 
%distribution has two more variables. Using the standard NNPDF3.1 set without 
%$NISR gives compatible results more quickly. 

Earlier phenomenological studies with KKMChh~\cite{kkmchh2} 
followed this approach out of necessity. The fact 
that NISR with the NNPDF LuxQED PDF set gives compatible results confirms
the validity of that approach for NNPDF sets, which is also the one followed  
in the recent calculations for the ATLAS single-photon 
events presented at this conference~\cite{atlas:1gamma}.

%That these two approaches should have agreed for studies focused near
%the $Z$ pole is neither \emph{a priori} obvious nor universal. Tests with other
%PDF sets having QED-corrected versions show that MMHT2015~\cite{mmht} also has 
%a ratio of QED-corrected to normal PDF luminosity that closely matches the 
%KKMChh ISR factor near the $Z$ pole, while CT18~\cite{ct18} does not, for 
%either the CT18qed or CT18lux variants. Using an ordinary MMHT PDF set without
%NISR should also give compatible ISR results to a QED-corrected set with NISR. 
%We have not yet calculated MC distributions with CT18, but would not expect as
%close an agreement between the two methods in that case.

We expect to publish more details on the effects of exponentiated ISR
on quark distributions and the implementation of NISR in KKMChh 
shortly.~\cite{to-appear} We also envisage updating the {\tt photos} and 
{\tt tauola} libraries used by KKMChh and KKMCee; see Ref.~\cite{tauola} and 
references therein.

\acknowledgments

We acknowledge computing support from the Institute of Nuclear Physics IFJ-PAN, 
Krakow, and The Citadel. This talk was supported in part by a grant from the 
Citadel Foundation.  S. J. acknowledges funding from the European Union's
Horizon 2020 research and innovation program under grant agreement no.\ 951754
and support from the National Science Centre, Poland, grant 
no.\  2019/34/E/ST2/00457. Z. W. acknowledges funding from the Polish 
National Science Centre under decision DEC-2017/27/B/ST2/01391.

\end{document}